\documentclass[preprint,amsmath,amssymb,pre,longbibliography,superscriptaddress,longbibliography]{revtex4-1}
\usepackage{color,longtable}
\usepackage[dvips]{graphicx}
\usepackage{amssymb,amsfonts,amsmath}
\begin{document}
\title{
Geometric control of powder jet dynamics and energy dissipation
}
\author{Kazuya U. Kobayashi}
\affiliation{%
Department of Mechanical Engineering, Nippon Institute of Technology, 4-1 Gakuendai, Miyashiro-machi, Minamisaitama-gun, Saitama, 345-8501, Japan
}%
\author{Komei Jinbo}
\affiliation{%
Department of Mechanical Engineering, Nippon Institute of Technology, 4-1 Gakuendai, Miyashiro-machi, Minamisaitama-gun, Saitama, 345-8501, Japan
}%
\author{Riku Kodama}
\affiliation{%
Department of Mechanical Engineering, Nippon Institute of Technology, 4-1 Gakuendai, Miyashiro-machi, Minamisaitama-gun, Saitama, 345-8501, Japan
}%
\author{Masakazu Muto}
\affiliation{%
Department of Electrical and Mechanical Engineering, Nagoya Institute of Technology, Gokiso, Showa-ku, Nagoya-city, Aichi, 466-8555, Japan
}%
\author{Rei Kurita}%
\affiliation{%
Department of Physics, Tokyo Metropolitan University, 1-1 Minami-Osawa, Hachioji-city, Tokyo, 192-0397, Japan
}%
\date{\today}

\begin{abstract}
Applying an impulsive force to a powder layer shaped with a concave surface generates a sharp powder jet. This phenomenon has been proposed as a method for evaluating the flowability of powders from small amount of samples. In this study, we systematically varied the radius of the initial concave shape as a controllable parameter and quantitatively examined the resulting jet dynamics, focusing on ejection velocity and maximum height.
Our high-speed observations revealed that increasing the concave radius led to broader jets with significantly reduced velocity and maximum height. These dynamic quantities followed a scaling relation with drop height, while the scaling coefficient decreased with the concave radius, indicating that the surface geometry directly governs the extent of energy dissipation. Furthermore, a minimal mechanical model incorporating the sliding distance and velocity squared type dissipation of the powder flow reproduces the observed linear dependence of the jet height on the concave radius. 
These findings establish powder jets as a sensitive probe of dissipation in dynamic powder flow and provide a quantitative framework for comparing powder specific interactions such as humidity, particle size and particle shape.
\end{abstract}
\maketitle

\section{Introduction}
It is well known that powder materials exhibit complex dynamics, combining solid-like and liquid-like behaviors due to interparticle interactions such as collisions, friction, adhesion, and rigid-body contacts~\cite{Duran1999, Andreotti2013, Partelli2014, Vescovi2016, Zheng2024, Sharma2025}. These behaviors span a broad range of applications, from industrial operations like mixing and transportation to natural phenomena including landslides, volcanic eruptions, and planetary impact events~\cite{Campbell2006, Jop2006, Bridgwater2012, Jerolmack2019, Kobayashi2022, Rauter2022, Tahmasebi2023, Man2025}. The macroscopic responses of powders such as flow, jetting, and agglomerate collapse arise from how external forces are transmitted through particle assemblies and how energy is dissipated via inelastic collisions and friction~\cite{Silbert2001, Jop2006, Silbert2007, Degiuli2017, Katsuragi2016, Sun2023}.

Conventional methods for evaluating powder flowability, including angle of repose and powder rheometry, typically require large sample volumes and are susceptible to boundary and preparation conditions~\cite{Podczeck1997, Freeman2007, Hare2015, Wang2016, Shi2020, Tharanon2024}.
In contrast, powder jetting has been proposed as a small-scale, portable method for probing dynamic flow behavior in powders~\cite{Kobayashi2025-2}. 
Previous studies have demonstrated that when an impact is applied to a powder bed containing a concave-shaped cavity, a thin jet emerges vertically from the powder surface~\cite{Kobayashi2024}. These jets are highly sensitive to subtle interparticle forces, with the maximum jet height showing a marked dependence on ambient humidity~\cite{Kobayashi2025-2}.

While humidity-dependent interactions have been studied as a dominant factor in jet behavior, the dynamics of powder jets also reflect the underlying flow processes preceding jet formation. When an impact is applied, the input energy is transmitted into the powder bed, triggering particle motion along the inclined concave surface. This induces a transient sliding flow, during which particles converge toward the center of the cavity, leading to the accumulation of kinetic energy and subsequent jet formation~\cite{Meer2017, Huang2020}. Despite its central role, this pre-jet process has not been quantitatively characterized in detail.
During this sliding stage, a significant portion of the input energy is dissipated through interparticle interactions. The mode and magnitude of this energy dissipation are expected to be closely related to the flowability of the powder, yet remain largely unexplored. Since the jet dynamics depends on the energy carried by the particle flow, dissipation during the pre-jet phase becomes a key factor. In this context, the geometry of the initial cavity, particularly its radius, governs the distance and pattern of particle motion, providing a direct handle to control the extent of energy dissipation. This motivates a systematic investigation of how dissipation emerges from the underlying flow and how it governs the resulting jet dynamics.

In this study, we investigate how energy dissipation emerges during the pre-jet flow and how it governs the resulting jet dynamics. We systematically varied the cavity radius and performed high-speed imaging to quantify the resulting jet velocity, maximum height, and jet diameter. We found that increasing the cavity radius enlarged the jet diameter but significantly reduced both the jet speed and height, indicating enhanced energy dissipation during the flow process. By constructing a simple model that incorporates both energy accumulation and dissipation, we establish a generalized framework for the jet dynamics and show that the dissipation scales with the square of the sliding velocity in this glass-bead system.

\section{Materials and Experimental Methods}
\begin{figure}[t]
\begin{center}
\includegraphics[width=140mm]{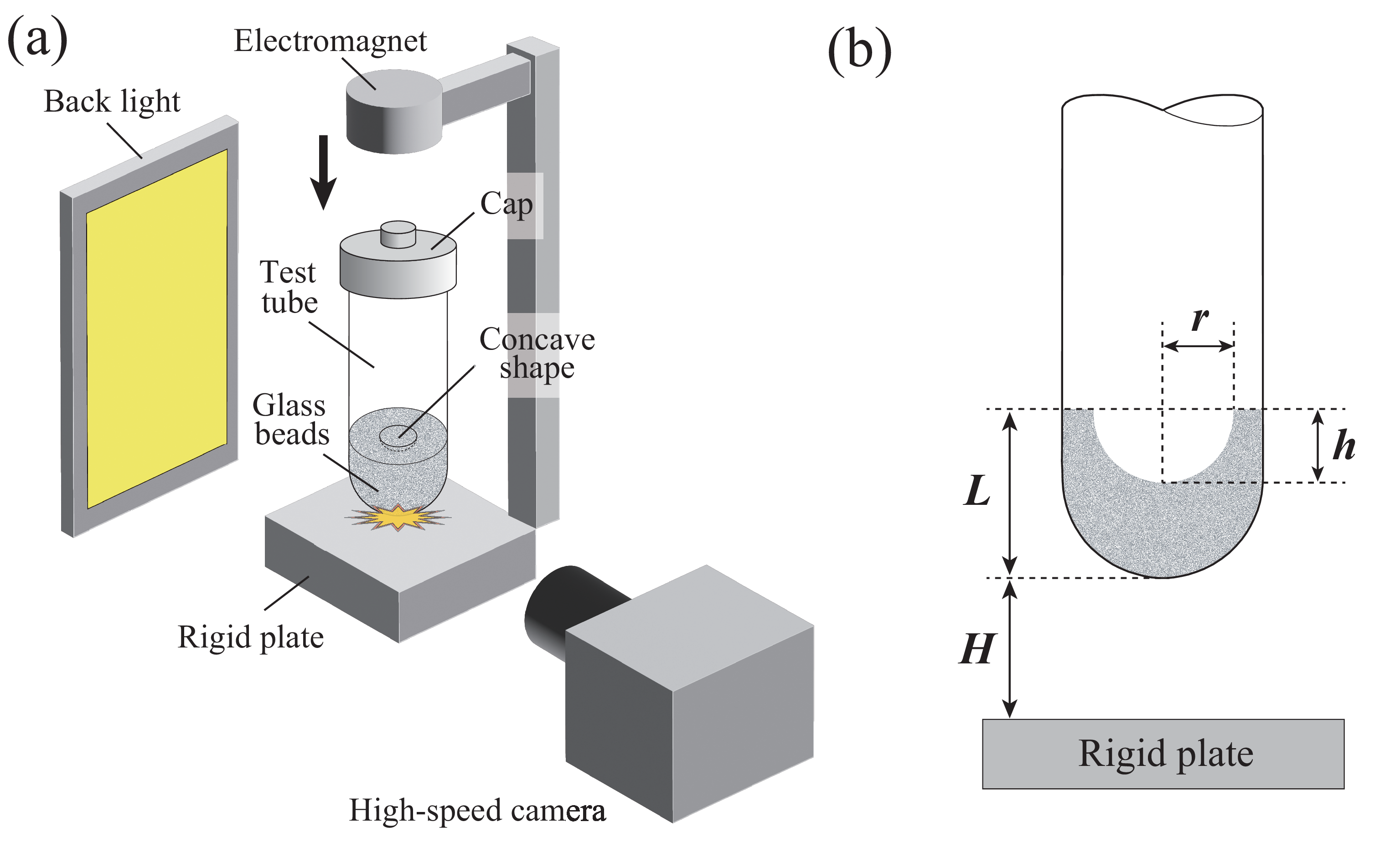}
\end{center}
\caption{
Schematic illustration of the experimental setup and definitions of the key parameters.
(a) Overview of the apparatus used to generate powder jets by impulsive loading.
(b) Definition of the concave geometry formed on the powder surface and the parameters
that govern the jet dynamics, including the concave radius $r$, depth $h$, filling height $L$,
and drop height $H$.  
These geometric parameters determine the surface flow distance and thereby the energy
dissipation prior to jet ejection.
}
\label{setup}
\end{figure}

A schematic of the experimental setup is shown in Fig. \ref{setup}(a).
The test tube was made of borosilicate glass, with an outer diameter of 30~mm, an inner diameter of 27~mm, and a length of 200~mm (Nichidenrika-Glass Co., Ltd.).
Glass beads with a uniform spherical shape and smooth surface (Fuji Manufacturing Co.) were used as the packing material.
Compared with powders of irregular shape, the use of spherical beads minimizes the influence of shape dispersion~\cite{Hafez2021}.
According to the manufacturer’s specifications, glass beads with a particle size range of 38--53~$\mu$m (hereafter referred to as 45~$\mu$m as the representative diameter) and a density of 2.5~g/cm$^3$ were used in this experiment.
As shown in Fig. \ref{setup}(b), the glass beads were filled into the test tube from the bottom to a height of $L$ = 25~mm.
To ensure that $L$ remained constant despite variations in the initial packing fraction $\phi$, the powder height was adjusted in advance by tapping the test tube several times.
The initial packing fraction was approximately $\phi$ = 0.55.
To create a focused jet, a rod with a hemispherical tip and radii $r$ = 5, 7.5, 10.5, and 12~mm was used to form a concave surface on the powder layer.
The depth $h$ of the concave shape was adjusted such that $h = r$, ensuring that the concave surfaces were geometrically similar for different values of $r$ (Fig. \ref{setup}(b)).

The test tube was sealed with a silicone cap equipped with a metal hook for electromagnetic fixation.
It was then suspended vertically using an electromagnet and brought to rest.
When the electromagnet was switched off, the test tube underwent free fall and collided with a pre-installed metal floor, generating a powder jet upon impact.
The drop height $H$ was varied in the range of 10--110~mm.
The experimental configuration is similar to that used in previous studies on liquid jets generated by impact~\cite{Antkowiak2007, Kiyama2016, Gordillo2020}. 
The jet formation process was recorded using a high-speed camera (FASTCAM SA5, Photron Co.) at a frame rate of 30{,}000 frames per second (fps).
Each experiment was repeated five times under identical conditions, and the mean values and standard deviations are reported as representative values and error bars, respectively.

Measurements of the tilt of the test tube during free fall showed that, in all experiments reported here, the inclination angle was within 5 degrees, and the resulting variation in jet velocity was within 10~\%.
These results confirm that the influence of tube tilt on the jet behavior was negligible, demonstrating the reproducibility of the experimental method.
To ensure that the powder particles reached equilibrium at a relative humidity of 70~\%, the samples were stored in a humidity-controlled chamber at 70~\% relative humidity for 24~h prior to the experiment.
The temperature was maintained at 23~$^\circ$C.
Following a previous study~\cite{Kobayashi2025-2}, the mass of the powder was measured at regular intervals under conditions of 70~\% relative humidity and 23~$^\circ$C to confirm humidity equilibrium.
All samples exhibited a mass change $\Delta m$ of less than 0.001~g over more than 24~h, indicating that a standing time of 24~h was sufficient to achieve equilibrium.
All experiments were conducted in a clean booth with minimal air inflow from outside, and the ambient humidity was controlled using a humidifier to match that of the humidity-controlled chamber.

\section{Results}
\subsection{Jet Generation Behavior in Response to Variations in Concave Surface Shape}
\begin{figure}[t]
\begin{center}
\includegraphics[width=150mm]{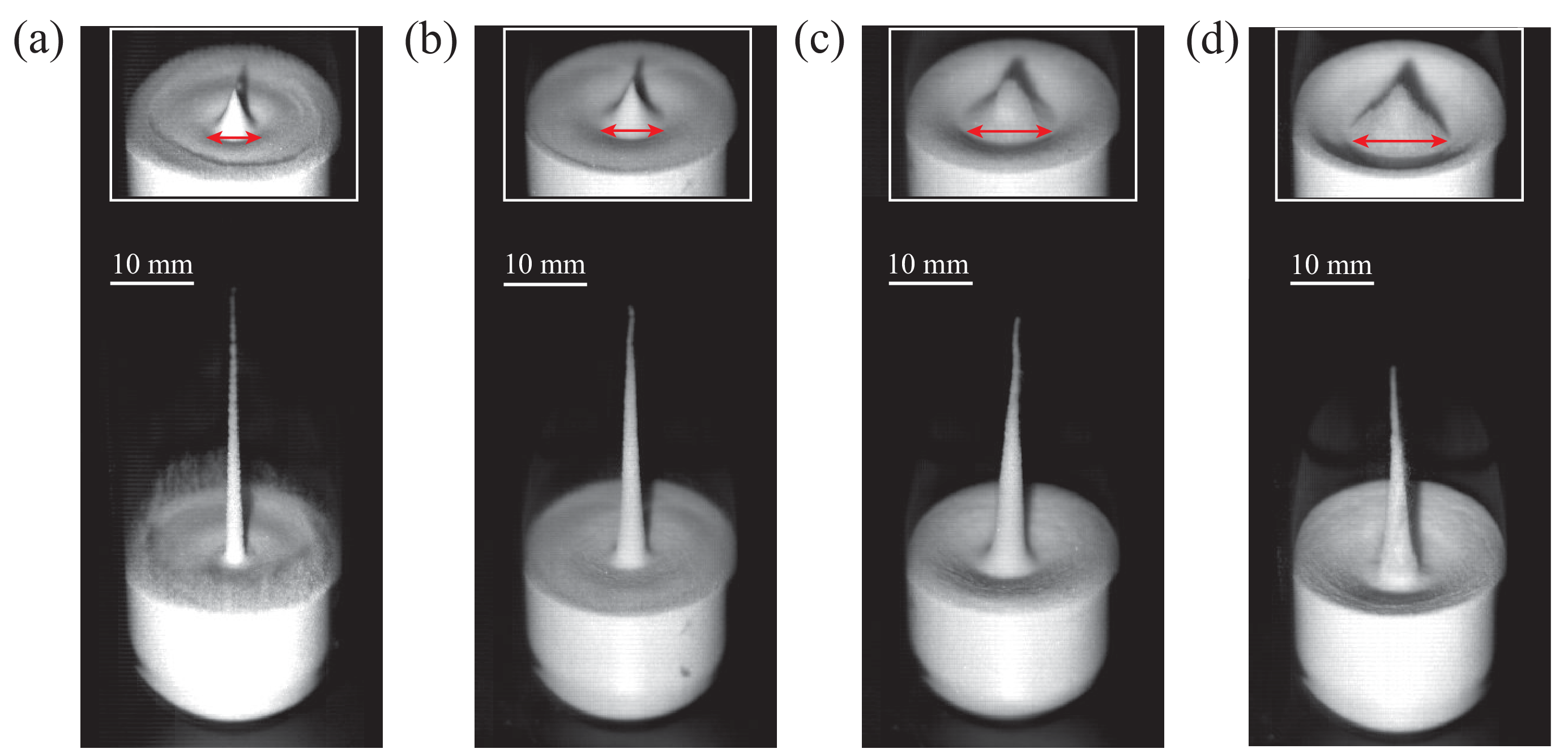}
\end{center}
\caption{
Diagonal view snapshots of powder jet formation for different initial concave radii $r$.
Panels (a) to (d) correspond to $r$ = 5, 7.5, 10.5 and 12 mm.  
Each sequence shows the jet evolution at 30 ms.
The insets display the early stage of jet emergence at 6 ms.  
A clear geometric effect is observed: as $r$ increases, the jet becomes wider and its
ejection height becomes lower.  
This demonstrates that the concave radius directly modulates the internal dissipation
during the pre jet flow.
The drop height is fixed at $H = 10$ mm.
}
\label{jet2}
\end{figure}

First, we investigated the characteristic behavior of jet generation in response to variations in the concave surface shape. When the test tube collided with the metal floor (at $t$ = 0~ms), the concave surface collapsed and the powder flowed toward the center, generating a vertically upward jet~\cite{Kobayashi2025-2}.
The jet shape was recorded from an oblique overhead angle using a high-speed camera, and representative images are shown in Fig.~\ref{jet2}. Panels (a), (b), (c), and (d) correspond to concave surface radii of $r$ = 5, 7.5, 10.5, and 12~mm, respectively. Each image was taken 30~ms after the impact of the test tube with the floor. The free-fall height was fixed at $H$ = 10~mm, and the relative humidity was maintained at 70~\%.
The insets in each panel show magnified views of the central region of the concave surface captured 6~ms after impact. Jets were observed for all values of $r$; however, as $r$ increased, the jet height decreased while the jet diameter increased.
To quantify these trends, we analyzed the dependence of maximum jet reach $L_{max}$ and the jet velocity $V_{jet}$ on $r$.

\subsection{Dependence of the Maximum Jet Reach $L_{\rm max}$ on Concave Surface Radii $r$}
\begin{figure}[t]
\begin{center}
\includegraphics[width=140mm]{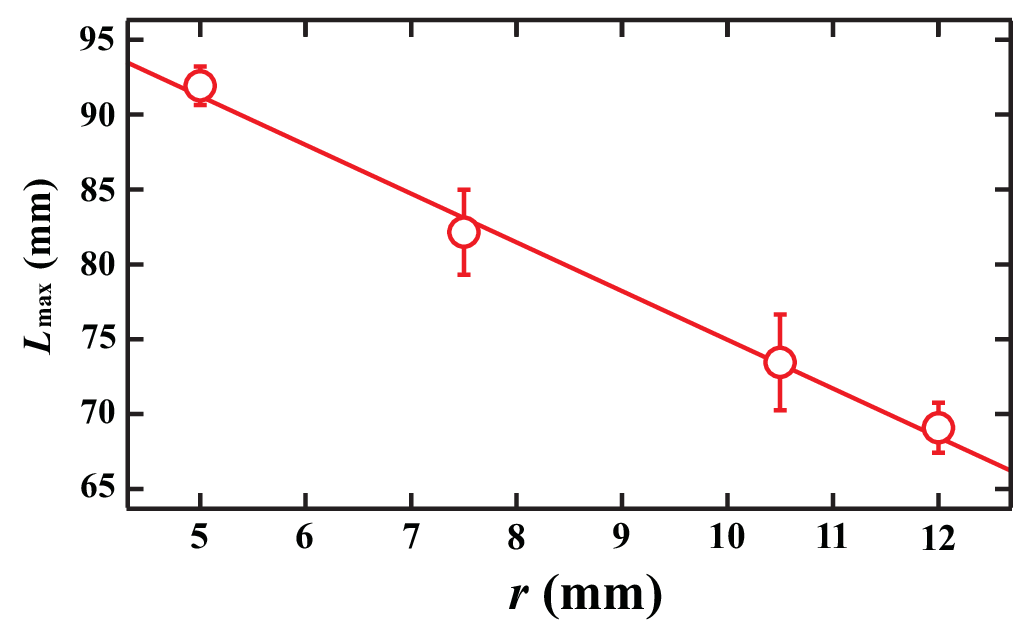}
\end{center}
\caption{
Maximum jet reach $L_{\rm max}$ as a function of the concave radius $r$ at fixed drop height $H = 10$ mm.  
The maximum height decreases approximately linearly with increasing $r$. 
}
\label{Lmax}
\end{figure}

We first examine the maximum jet reach $L_{\rm max}$, which can be measured with high accuracy and serves as a robust indicator of the jet dynamics.
Figure~\ref{Lmax} shows the dependence of $L_{\rm max}$ on the concave radius $r$ at a fixed drop height of $H = 10$~mm.
As $r$ increases, $L_{\rm max}$ decreases approximately linearly.
From the experimental data in Fig.~\ref{Lmax}, we obtain the linear relation
$L_{\rm max} = 0.107 - 3.23 r$~m, with a coefficient of determination exceeding 0.994.
Since $L_{\rm max}$ corresponds to the gravitational potential energy of the jet tip, this linear decrease indicates that the energy available for jet formation is reduced in proportion to $r$, reflecting enhanced energy dissipation during the pre-jet flow.
The intercept corresponds to the limit of $r \to 0$, where dissipation becomes negligible and thus represents the maximum energy accumulation in the absence of dissipation.
A quantitative theoretical description of the jet formation process will be discussed in Sec.~IV.

\subsection{Dependence of Jet Velocity $V_{\rm jet}$ on Concave Surface Radii $r$}
\begin{figure}[t]
\begin{center}
\includegraphics[width=140mm]{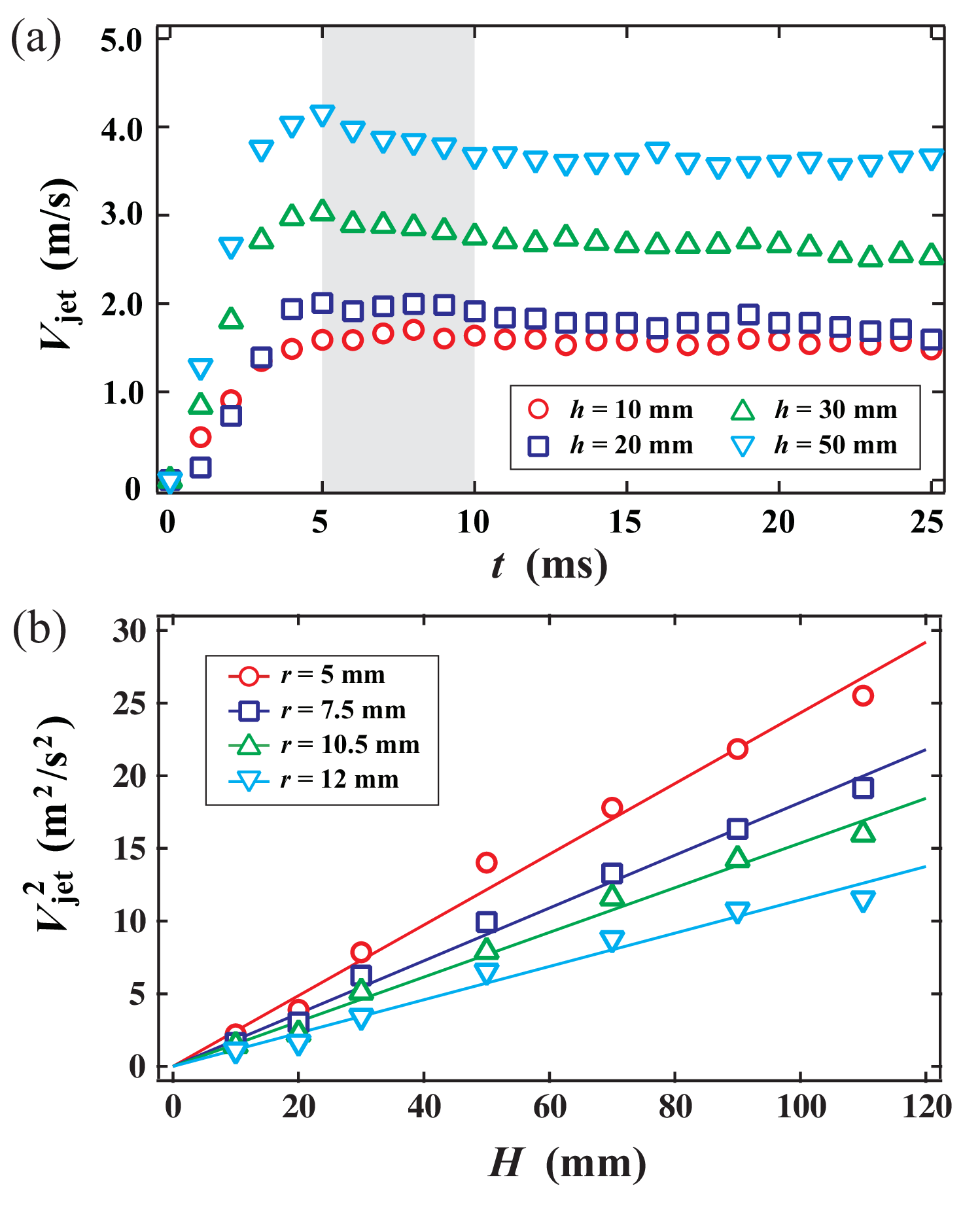}
\end{center}
\caption{
(a) Time evolution of the jet velocity. 
In the early stage, particles accelerate as they gain energy while sliding along the concave surface. 
The velocity becomes nearly constant in the time interval from 5 to 10 ms, corresponding to the period when the jet reaches its maximum speed.
(b) Squared jet velocity, $V_{\rm jet}^2$, as a function of the drop height $H$ for concave surface radii of $r$ = 5, 7.5, 10.5 and 12 mm.  
For all values of $r$, $V_{\rm jet}^2$ is proportional to $H$, indicating that the kinetic energy of the jet scales linearly with the initial potential energy.
}
\label{vmean}
\end{figure}

To quantitatively discuss energy dissipation in the powder flow, we analyze the jet formation dynamics by measuring the time evolution of the jet velocity (Fig.~\ref{vmean}(a)).
In the early stage, particles accelerate as energy is accumulated through sliding along the concave surface.
The velocity becomes nearly constant in the time interval from $t = 5$~ms to $10$~ms, corresponding to the period when the jet reaches its maximum speed.
We therefore define the representative jet velocity $V_{\rm jet}$ as
\[
V_{\rm jet} = \frac{z(\tau + \Delta t) - z(\tau)}{\Delta t},
\]
where $z(t)$ denotes the position of the jet tip measured from the powder surface at time $t$, and $\Delta t$ is the time interval.
We set $\tau = 5$~ms and $\Delta t = 5$~ms.
After this stage, the jet decelerates with an acceleration approximately equal to $-g$~\cite{Kobayashi2025-2}, indicating that no additional upward impulse acts on the jet tip.

Next, to examine the role of energy dissipation, we investigate the relationship between the initial gravitational potential energy and the kinetic energy of the jet tip.
Figure~\ref{vmean}(b) shows the squared jet velocity $V_{\rm jet}^2$ as a function of the drop height $H$ for concave surface radii of $r = 5$, $7.5$, $10.5$, and $12$~mm.
For all values of $r$, $V_{\rm jet}^2$ is proportional to $H$, indicating that the kinetic energy of the jet scales linearly with the initial potential energy.
This scaling is not trivial in the presence of dissipation during sliding along the concave surface.
As will be discussed later using a generalized theoretical framework, this relation holds only when the dissipation scales with the square of the sliding velocity.
The dissipation in this system is likely dominated by internal flow resistance within the sliding powder layer, rather than simple Coulomb-like interparticle friction. In high-velocity granular or particle-laden flows such as the present glass-bead system, the resistive force is often described as scaling with the square of the flow velocity, reflecting inertial and collisional interactions among particles in dense granular flows.
In addition, $V_{\rm jet}^2$ decreases monotonically with increasing $r$, consistent with the trend observed in $L_{\rm max}$.

\begin{figure}[t]
\begin{center}
\includegraphics[width=140mm]{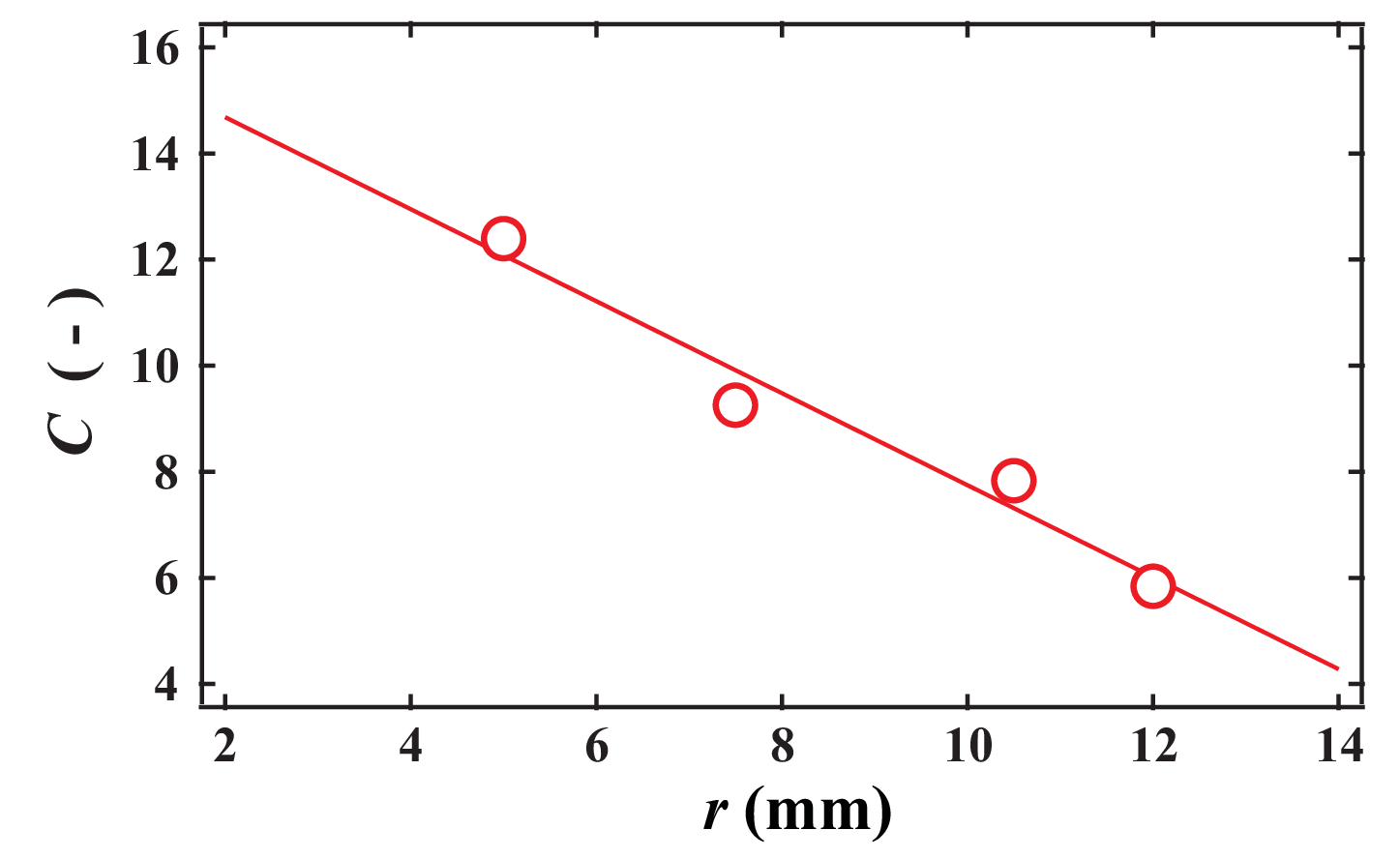}
\end{center}
\caption{
Dependence of the fitting coefficient $C$ on the concave radius $r$. 
The value of $C$ decreases linearly with increasing $r$, indicating that the efficiency of momentum transfer to the jet is reduced as the concave geometry becomes wider. 
This trend suggests that the surface geometry strongly influences the energy dissipation characteristics of the powder flow.
}
\label{C_r}
\end{figure}

The dependence of the linear fitting coefficient $C$ on the concave radius $r$ is summarized in Fig.~\ref{C_r}.
The value of $C$ decreases linearly with increasing $r$, indicating that the efficiency of momentum transfer to the jet becomes lower as the concave shape becomes wider.
This result suggests that the surface geometry strongly affects the dissipation characteristics of the powder flow.
Even though the total powder volume, ambient humidity (70~\% RH), and impact energy (set by $H$) were kept constant, the observed changes in $V_{\rm jet}$ as a function of $r$ reflect intrinsic differences in the flow dynamics along the inclined surface.
Since the powder jet is formed by the collapse of the concave surface and the subsequent inward flow of particles, the surface flow behavior preceding jet formation plays a central role in determining the jet velocity~\cite{Katsuragi2016, Kobayashi2025-2}.
Figure~\ref{C_r} shows that $C$ varies linearly with $r$ as $C = 16.4 - 8.64 \times 10^{2} r$.

As discussed in connection with Fig.~\ref{vmean}(a), once the jet tip reaches this stage, it decelerates with an acceleration close to $-g$, indicating that no additional upward driving force acts from below. The subsequent rise can therefore be regarded as approximately ballistic, and the above relation is used as a leading-order approximation based on the conversion of the jet-tip kinetic energy into gravitational potential energy.
Then,
\[
\frac{1}{2} m_{\rm tip} V_{\rm jet}^2 = m_{\rm tip} g L_{\rm max},
\]
where $m_{\rm tip}$ is the mass of the jet tip, we obtain the relation
$L_{\rm max} = \frac{V_{\rm jet}^2}{2g} = C H$. 
Using the relation for $L_{\rm max}$ obtained independently, we find $C = 10.7 - 3.23 \times 10^{2} r$.
Although the coefficients differ slightly, both estimates show the same functional dependence on $r$ and are consistent in order of magnitude.
The slight difference in the coefficients may arise from differences in the evaluation procedures.

\section{Discussion}
Here, we propose an $r$-dependent model of jet dynamics based on classical mechanics.
We consider the jet formation process in a time-sequential manner.
The velocity of the tube just before impact is determined by the initial gravitational potential energy as
\[
V_{\rm in} = -\sqrt{2gH}.
\]
After collision with the floor, the rebound velocity is given by
\[
V_{\rm out} = e \sqrt{2gH},
\]
where $e$ is the restitution coefficient.
Following the impact, particles on the concave surface begin to slide.
Assuming a uniform sliding layer with thickness $\xi$, the sliding mass is expressed as
\[
m_{\rm slide} = \int_S \rho dS \cdot \xi.
\]
We define the spatially averaged sliding velocity, $V_{\mathrm{slide}}$, as a characteristic velocity scale at the onset of the avalanche-like sliding motion immediately after impact, before dissipation has accumulated substantially. Since this initial sliding is driven by the rebound and the concave geometries are mutually similar, $V_{\mathrm{slide}}$ is not expected to depend on $r$.
During this sliding process, energy dissipation occurs due to interparticle interactions.
The dissipation mechanism likely depends on the flow properties of the powder system. In powders with weak interparticle interactions, flow may occur more readily, whereas in cohesive powders or systems composed of irregular particles, resistance may become more pronounced because of attractive interactions and structural dissipation associated with compaction, force-chain rearrangements, and shear localization.
We therefore assume that the dissipation scales as $V_{\rm slide}^\beta$, where $\beta$ is a system-dependent exponent.
The dissipated energy is written as
\[
E_{\rm dis} = \epsilon \frac{\pi}{2} r m_{\rm slide} V_{\rm slide}^\beta,
\]
where $\epsilon$ is a dissipation coefficient per unit mass.

The energy balance between the sliding state and the jet formation is then given by
\[
\frac{1}{2} m_{\rm slide} V_{\rm slide}^2 - E_{\rm dis}
= \frac{1}{2} m_{\rm jet} V_{\rm jet}^2,
\]
where $m_{\rm jet}$ is the mass of the jet described as $m_{\rm jet} = \int_{jet} \rho dV$. 
We assume that the sliding velocity is proportional to the rebound velocity as
\[
V_{\rm slide} = c_1 V_{\rm out} = c_1 e \sqrt{2gH},
\]
where $c_1$ is a proportionality constant.
We also define the mass ratio as
\[
\alpha \equiv \frac{m_{\rm slide}}{m_{\rm jet}}.
\]
Substituting these relations into the energy balance equation yields
\[
V_{\rm jet}^2
= \alpha c_1^2 e^2 (2gH)
- \epsilon \pi \alpha c_1^\beta e^\beta (2gH)^{\beta/2} r.
\]
This expression shows that the scaling behavior of $V_{\rm jet}^2$ with respect to $H$ depends on the exponent $\beta$.
In the present experiments, we observed that $V_{\rm jet}^2 \propto H$ holds over a wide range of $H$.
This requirement is satisfied only when $\beta = 2$, indicating that the dissipation scales with the square of the sliding velocity.
In the present experiments, the observed scaling $V_{\mathrm{jet}}^{2} \propto H$ requires $\beta = 2$, indicating velocity-squared dissipation during the pre-jet flow. 
This behavior is consistent, at the scaling level, with inertial granular-flow descriptions such as Bagnold-type~\cite{Bagnold1954}, $\mu(I)$-based frameworks~\cite{GDR2004,Jop2006}, or Voellmy-type formulations~\cite{Bartelt1999,Bartelt2006}. In the present study, however, the dissipation term is introduced phenomenologically to describe the observed jet-formation behavior, rather than as a direct application of any one of these theories.
If the dissipation mechanism differs from that in the present system, for example because of stronger interparticle interactions or lower flow velocities, $\beta < 2$ may be expected. 
In such cases, the linear relation between $V_{\rm jet}^2$ and $H$ would break down, particularly at smaller values of $H$. 
In addition, since $L_{\max}$ is proportional to $V_{\mathrm{jet}}^{2}$ through gravitational conversion, the same argument also applies to $L_{\max}$. 

For $\beta = 2$, the expression reduces to
\[
V_{\rm jet}^2 = (2gH)\left( \alpha c_1^2 e^2 - \epsilon \pi \alpha c_1^2 e^2 r \right),
\]
which leads to the coefficient
\[
C = \alpha c_1^2 e^2 - \epsilon \pi \alpha c_1^2 e^2 r.
\]
The first term represents the efficiency of energy accumulation, characterized by the mass ratio $\alpha$, while the second term represents energy dissipation through the coefficient $\epsilon$.
From the experimental observations, we estimate $c_1 \sim 2$ and $e \sim 0.8$.
Using the $r$-dependence of $L_{\rm max}$, we obtain $\alpha = 3.9$ and $\epsilon = 9.6$.

The linear relation between $L_{\rm max}$ and $r$ obtained in the present study is expected to hold only within a regime where avalanche-like sliding develops clearly and the associated dissipation is well described by $\beta = 2$. This condition is most likely satisfied when the impact is sufficiently strong that the pre-jet flow remains inertia-dominated. By contrast, if the impact energy is too small, avalanche-like sliding may not be fully developed, or the sliding velocity may become sufficiently low that the effective dissipation deviates from the velocity-squared form. In such cases, the proportionality $V_{\mathrm{jet}}^2 \propto H$ and the resulting linear dependence on $r$ would no longer be expected to hold. In addition, for sufficiently large $r$, the longer sliding path may cause substantial velocity reduction during motion, which could alter the effective dissipation behavior and lead to nonlinear deviations. A further limitation arises when the particle size becomes comparable to the geometric scale of the concave surface, in which case the continuum-like description of the sliding layer may break down and jet formation itself may become difficult.

\section{Conclusion}
This study demonstrated that the geometry of a concave surface formed on a powder layer provides direct control over the dynamics of powder jets generated by impulsive loading.  
Increasing the concave radius produced wider jets with lower ejection velocities and smaller maximum heights, and these trends were fully captured by the gravitational scaling with respect to the drop height.  
The concave geometry therefore governs the amount of internal dissipation that develops during the surface flow preceding jet formation.

A simple mechanical framework based on the balance between energy accumulation and dissipation is proposed.  
Within this framework, the observed gravitational scaling of the jet dynamics is explained by velocity-squared dissipation during sliding.  
Furthermore, this approach enables quantitative estimation of both the energy accumulation ratio and the magnitude of dissipation.

More generally, the geometric-control framework introduced here provides a predictive basis for understanding powder motion under impulsive loading through the interplay between geometry and dissipation. The generalized formulation further suggests that the same framework may be applicable to other sufficiently flowable particulate or soft matter systems that undergo avalanche-like motion, with material-dependent dissipation reflected in the effective exponent $\beta$. The present approach therefore offers a physically interpretable basis for probing dissipation mechanisms beyond the specific system studied here.

\section*{Author contributions}
Kazuya U. Kobayashi: Writing - review $\&$ editing; Writing - original draft; Validation; Project administration; Methodology; Investigation;
Funding acquisition; Conceptualization. 
Komei Jinbo: Writing - review $\&$ editing; Writing - original draft; Validation, Investigation.
Riku Kodama: Writing - review $\&$ editing, Writing - original draft, Validation; Investigation.
Masakazu Muto: Writing - review $\&$ editing; Writing - original draft; Validation; Investigation. 
Rei Kurita: Writing - review $\&$ editing; Writing - original draft; Validation; Methodology; Investigation; Conceptualization.

\section*{Conflicts of interest}
There are no conflicts of interest to declare.

\section*{Data availability}
All data supporting the findings of this study are included within the article.

\section*{Acknowledgements}
This study was partly supported by a Grant-in-Aid for Early-Career Scientists (Grant No. 24K17024) from the Japan Society for the Promotion of Science (JSPS) and a Suzuki Foundation. R.K. was supported by a Grant-in-Aid for Scientific Research (B) (Grant No. 26K00676) from the JSPS.

\section*{Correspondence}
Correspondence and requests for materials should be addressed to K. U. K. (kobayashi.kazuya@nit.ac.jp) and R. K. (kurita@tmu.ac.jp).

\clearpage

\end{document}